\documentclass[dvips]{article}
\usepackage{supertabular,lscape,epsfig}
\usepackage{amssymb}
\usepackage{amsmath}

\DeclareSymbolFont{ppa}{OT1}{ppl}{m}{it}
\DeclareMathSymbol{\vv}{\mathalpha}{ppa}{'166}

\begin{document}

\newcommand{\dd}{\,{\rm d}}
\newcommand{\ie}{{\it i.e.},\,}
\newcommand{\etal}{{\it et al.\ }}
\newcommand{\eg}{{\it e.g.},\,}
\newcommand{\cf}{{\it cf.\ }}
\newcommand{\vs}{{\it vs.\ }}
\newcommand{\zdot}{\makebox[0pt][l]{.}}
\newcommand{\up}[1]{\ifmmode^{\rm #1}\else$^{\rm #1}$\fi}
\newcommand{\dn}[1]{\ifmmode_{\rm #1}\else$_{\rm #1}$\fi}
\newcommand{\upd}{\up{d}}
\newcommand{\uph}{\up{h}}
\newcommand{\upm}{\up{m}}
\newcommand{\ups}{\up{s}}
\newcommand{\arcd}{\ifmmode^{\circ}\else$^{\circ}$\fi}
\newcommand{\arcm}{\ifmmode{'}\else$'$\fi}
\newcommand{\arcs}{\ifmmode{''}\else$''$\fi}
\newcommand{\MS}{{\rm M}\ifmmode_{\odot}\else$_{\odot}$\fi}
\newcommand{\RS}{{\rm R}\ifmmode_{\odot}\else$_{\odot}$\fi}
\newcommand{\LS}{{\rm L}\ifmmode_{\odot}\else$_{\odot}$\fi}

\newcommand{\Abstract}[2]{{\footnotesize\begin{center}ABSTRACT\end{center}
\vspace{1mm}\par#1\par
\noindent
{~}{\it #2}}}

\newcommand{\TabCap}[2]{\begin{center}\parbox[t]{#1}{\begin{center}
 \small {\spaceskip 2pt plus 1pt minus 1pt T a b l e}
 \refstepcounter{table}\thetable \\[2mm]
 \footnotesize #2 \end{center}}\end{center}}

\newcommand{\TableSep}[2]{\begin{table}[p]\vspace{#1}
\TabCap{#2}\end{table}}

\newcommand{\FigCap}[1]{\footnotesize\par\noindent Fig.\ %
 \refstepcounter{figure}\thefigure. #1\par}

\newcommand{\TableFont}{\footnotesize}
\newcommand{\TableFontIt}{\ttit}
\newcommand{\SetTableFont}[1]{\renewcommand{\TableFont}{#1}}

\newcommand{\MakeTable}[4]{\begin{table}[htb]\TabCap{#2}{#3}
 \begin{center} \TableFont \begin{tabular}{#1} #4 
 \end{tabular}\end{center}\end{table}}

\newcommand{\MakeTableSep}[4]{\begin{table}[p]\TabCap{#2}{#3}
 \begin{center} \TableFont \begin{tabular}{#1} #4 
 \end{tabular}\end{center}\end{table}}

\newenvironment{references}%
{
\footnotesize \frenchspacing
\renewcommand{\thesection}{}
\renewcommand{\in}{{\rm in }}
\renewcommand{\AA}{Astron.\ Astrophys.}
\newcommand{\AAS}{Astron.~Astrophys.~Suppl.~Ser.}
\newcommand{\ApJ}{Astrophys.\ J.}
\newcommand{\ApJS}{Astrophys.\ J.~Suppl.~Ser.}
\newcommand{\ApJL}{Astrophys.\ J.~Letters}
\newcommand{\AJ}{Astron.\ J.}
\newcommand{\IBVS}{IBVS}
\newcommand{\PASP}{P.A.S.P.}
\newcommand{\Acta}{Acta Astron.}
\newcommand{\MNRAS}{MNRAS}
\renewcommand{\and}{{\rm and }}
\section{{\rm REFERENCES}}
\sloppy \hyphenpenalty10000
\begin{list}{}{\leftmargin1cm\listparindent-1cm
\itemindent\listparindent\parsep0pt\itemsep0pt}}%
{\end{list}\vspace{2mm}}

\def\TYLDA{~}
\newlength{\DW}
\settowidth{\DW}{0}
\newcommand{\dw}{\hspace{\DW}}

\newcommand{\refitem}[5]{\item[]{#1} #2%
\def\REFARG{#3}\ifx\REFARG\TYLDA\else, {\it#3}\fi
\def\REFARG{#4}\ifx\REFARG\TYLDA\else, {\bf#4}\fi
\def\REFARG{#5}\ifx\REFARG\TYLDA\else, {#5}\fi.}

\newcommand{\Section}[1]{\section{#1}}
\newcommand{\Subsection}[1]{\subsection{#1}}
\newcommand{\Acknow}[1]{\par\vspace{5mm}{\bf Acknowledgements.} #1}
\pagestyle{myheadings}

\def\thefootnote{\fnsymbol{footnote}}
\begin{center}
{\Large\bf The Optical Gravitational Lensing Experiment.\\
\vskip3pt
Cepheids in the Magellanic Clouds.\\
\vskip3pt
VI. Double-Mode Cepheids\\
\vskip3pt
in the Large Magellanic Cloud\footnote{Based on observations obtained with the 
1.3~m Warsaw telescope at the Las Campanas Observatory of the Carnegie 
Institution of Washington.}} 
\vskip1cm
{\bf
I.~~S~o~s~z~y~{\'n}~s~k~i$^1$,~~A.~~U~d~a~l~s~k~i$^1$,
~~M.~~S~z~y~m~a~{\'n}~s~k~i$^1$,~~M.~~K~u~b~i~a~k$^1$,
~~G.~~P~i~e~t~r~z~y~\'n~s~k~i$^{1,2}$,
~~P.~~W~o~\'z~n~i~a~k$^3$,~~ and~~K.~~\.Z~e~b~r~u~\'n$^1$}
\vskip3mm
{$^1$Warsaw University Observatory, Al.~Ujazdowskie~4, 00-478~Warszawa, Poland\\
e-mail: (soszynsk,udalski,msz,mk,pietrzyn,zebrun)@astrouw.edu.pl\\
$^2$ Universidad de Concepci{\'o}n, Departamento de Fisica,
Casilla 160--C, Concepci{\'o}n, Chile\\
$^3$ Princeton University Observatory, Princeton, NJ 08544-1001, USA\\
e-mail: wozniak@astro.princeton.edu}
\vskip5mm
\end{center}

\Abstract{We present a sample of 76 double-mode Cepheids detected in the 
4.5 square degree area in the central part of the LMC. 19 stars from the 
sample pulsate in the fundamental mode and the first overtone while 57 
objects are the first and second overtone pulsators. 

We analyze the period ratio of double-mode Cepheids and Fourier 
parameters of decomposition of the light curves of these objects. We 
also present location of different type Cepheids from the LMC in the 
color-magnitude diagram and show the distribution of their ${V-I}$ color 
indices.}

\Section{Introduction} 
Observations of double-mode Cepheids provide many information about 
evolution and structure of massive stars. Parameters of these stars that 
can be measured with high precision, like periods, period ratios and 
Fourier coefficients of light curve decomposition, can give us important 
information not only on the basic physical parameters of these stars but 
also on chemical composition or distance to them.

Only 14 double-mode Cepheids are known in the Galaxy (Pardo and Poretti 
1997). In this sample only one object, CO~Aur, pulsates  simultaneously
in the first and the second overtone (FO/SO) modes. The  remaining Galactic
double-mode Cepheids have the fundamental mode and the first overtone excited
(FU/FO). Cepheids pulsating in two radial modes are  difficult to
discover, because of the large number of datapoints  required to detect
and characterize them. Gravitational microlensing  surveys provide
unique observational material for search for double mode  Cepheids.
Regular, long term photometric observations of high stellar  density
regions of the sky, are ideal for examining variable stars,  including
double-mode Cepheids. 

Forty five double-mode Cepheids were identified in the Large Magellanic
Clouds  by the MACHO microlensing team (Alcock \etal 1995). This sample,
increased later to 75 objects (Alcock \etal 1999), allowed detailed
study of properties of these objects. Based on the Fourier decomposition of
FO/SO double-mode Cepheids light curves, Alcock \etal (1999) 
characterized properties of the second overtone pulsations, what led to 
detection of potential candidates for the single-mode second overtone 
Cepheids (Alcock \etal 1999, Udalski \etal 1999b). 

The second phase of the Optical Gravitational Lensing Experiment (OGLE-II) 
microlensing survey started in January 1997. Since then the Large and 
Small Magellanic Clouds have been observed regularly, practically on every 
clear night. Observations were made through the {\it BVI} filters, very 
closely reproducing the standard {\it BVI} system. More than three years 
of monitoring allowed to detect and characterize thousands of Cepheids 
in both Magellanic Clouds including large sample of double-mode 
Cepheids. 

In the previous papers of this series we presented analysis of 93 
double-mode Cepheids detected by the OGLE team in the SMC (Udalski \etal 
1999a), discovery of 13 candidates for Cepheids pulsating solely in the 
second overtone (Udalski \etal 1999b), analysis of the Period-Luminosity 
and Period-Luminosity-Color relations of Cepheids from the LMC and SMC 
(Udalski \etal 1999c) and Catalogs of Cepheids in the LMC (Udalski \etal 
1999d) and SMC (Udalski \etal 1999e). 

This paper, the final of this series, completes the sample of Cepheids 
detected in Magellanic Clouds during the second phase of the OGLE 
project. We present here 76 double-mode Cepheids detected in the LMC. 
Nineteen of these objects are stars pulsating in the fundamental mode 
and the first overtone, 57 are the first and the second overtone 
pulsators. Forty two double-mode Cepheids from our sample were
previously  detected by the MACHO team (Alcock \etal 1995, 1999), the
remaining  stars are detected for the first time. We provide all basic 
observational parameters of detected stars. Their photometry is 
available from the OGLE Internet archive. 

\Section{Observations}
All observations presented in this paper were carried out during the 
second phase of the OGLE experiment with the 1.3-m Warsaw telescope at 
the Las Campanas Observatory, Chile, which is operated by the Carnegie 
Institution of Washington. The telescope was equipped with the "first 
generation" camera with a SITe ${2048\times2048}$ CCD detector working 
in the drift-scan mode. The pixel size was 24~$\mu$m giving the 0.417 
arcsec/pixel scale. Observations of the LMC were performed in the "slow" 
reading mode of the CCD detector with the gain 3.8~e$^-$/ADU and readout 
noise about 5.4~e$^-$. Details of the instrumentation setup can be found 
in Udalski, Kubiak and Szyma{\'n}ski (1997). 

Observations of the LMC started on January~6, 1997. 11 driftscan fields, 
each covering ${14.2\times57}$ arcmins on the sky, were observed during the 
first months of 1997. Additional 10 fields were added in October 1997 
increasing the total observed area of the LMC to about 4.5 square 
degree. In this paper we present data collected up to the end of May 
2000. Observations were obtained in the standard {\it BVI}-bands with 
the majority of measurements made in the {\it I}-band. The reader is 
referred to Udalski \etal (2000) for more details about methods of data 
reduction, tests on quality of photometric data, astrometry, location of 
observed fields etc. 
\vspace*{9pt}
\Section{Selection of Double-Mode Cepheids}
\vspace*{5pt}
The search for Cepheids in the 21 OGLE LMC fields is described in detail 
in Udalski \etal (1999d). The variable stars classified as Cepheids
(about 1500 objects in total) were then searched for double-mode objects. 

The selection of double-mode Cepheids was performed using identical two 
stage algorithm as applied for the SMC double-mode Cepheid search (Udalski 
\etal 1999a). In the first stage the observations of each Cepheid 
candidate were folded with the detected period, the light curve was 
fitted by high order polynomial and subtracted. The residuals were then 
searched for additional periodic signal and, if detected, such a 
candidate was selected for further analysis. Then a histogram of the ratio 
of the shorter to the longer period of double-mode Cepheid 
candidates was constructed. It exhibited two clear sharp peaks 
corresponding to the ratio of the first overtone to the fundamental 
period, ${\approx0.72}$, and the second to the first overtone period, 
${\approx0.805}$, in good agreement with Alcock \etal (1995, 1999). The list 
of the double-mode Cepheid candidates from this search included stars 
having the period ratio within ${\pm0.02}$ from these values. 

The second, final search for double-mode Cepheids was performed using the 
{\sc Clean} algorithm of period determination (Roberts, Leh{\'a}r and 
Dreher 1987). All 1500 objects from the LMC Cepheid candidate list were 
analyzed with the {\sc Clean} period search algorithm. Having well 
established limits for the period ratio of double-mode Cepheids from the 
preliminary analysis, only those objects which exhibited suitable period 
ratio (${\pm0.015}$) between the highest peak in the power spectrum and 
one of the next four strongest peaks were further analyzed. The final 
list of the double-mode Cepheid candidates presented in this paper was 
obtained after careful visual inspection of the {\sc Clean} power 
spectra of each object. 

\begin{landscape}
\renewcommand{\arraystretch}{1.1}
\tabcolsep=5pt
\MakeTable{crccccccccccccl}{12.5cm}{FU/FO Double-Mode Cepheids in the LMC}
{\hline
\noalign{\vskip2pt}
Field   & Star No.& RA(J2000)                & DEC(J2000)                     & $P_{\rm FO}$ & $R_{21}^{\rm FO}$ & $\phi_{21}^{\rm FO}$ & $P_{\rm FU}$ & $R_{21}^{\rm FU}$ & $\phi_{21}^{\rm FU}$ & $P_{\rm FO}/P_{\rm FU}$ & $I$  & ${B-V}$ & ${V-I}$ & Remarks\\
        &         &                          &                                & [days]       &                   &                      & [days]       &                   &                      &                         & [mag]& [mag]   & [mag]   &\\
\noalign{\vskip2pt}
\hline
\noalign{\vskip2pt}
 LMC$\_$SC1 & 158021 &  5\uph33\upm39\zdot\ups53 & $-69\arcd54\arcm54\zdot\arcs8$ &  2.10084 &  0.159 &  4.920 &  2.93701 &  0.228 &  4.462 &   0.71530 & 15.373 &  0.495 &  0.705 &      M\\
 LMC$\_$SC2 & 143088 &  5\uph31\upm09\zdot\ups00 & $-70\arcd05\arcm14\zdot\arcs9$ &  2.45578 &  0.106 &  5.127 &  3.44679 &  0.209 &  4.401 &   0.71248 & 15.036 &  0.456 &  0.660 &      M\\
 LMC$\_$SC2 & 172961 &  5\uph31\upm00\zdot\ups05 & $-69\arcd49\arcm16\zdot\arcs3$ &  2.64048 &  0.102 &  5.126 &  3.68579 &  0.214 &  4.570 &   0.71639 & 15.002 &  0.514 &  0.731 &      M\\
 LMC$\_$SC3 &  88012 &  5\uph28\upm00\zdot\ups16 & $-69\arcd37\arcm20\zdot\arcs5$ &  1.12078 &  0.199 &  4.280 &  1.54689 &  0.180 &  4.476 &   0.72454 & 16.304 &  0.452 &  0.652 &      M\\
 LMC$\_$SC3 & 415237 &  5\uph29\upm36\zdot\ups16 & $-69\arcd40\arcm26\zdot\arcs8$ &  2.43755 &  0.091 &  5.254 &  3.40504 &  0.158 &  4.570 &   0.71587 & 15.226 &  0.550 &  0.749 &      M\\
 LMC$\_$SC4 & 436072 &  5\uph27\upm15\zdot\ups98 & $-69\arcd43\arcm43\zdot\arcs8$ &  2.44086 &  0.085 &  4.930 &  3.43369 &  0.250 &  4.499 &   0.71086 & 15.023 &  0.466 &  0.668 &      M\\
 LMC$\_$SC6 &      6 &  5\uph20\upm07\zdot\ups15 & $-70\arcd04\arcm09\zdot\arcs1$ &  3.45325 &  0.103 &  3.704 &  4.84127 &  0.165 &  4.591 &   0.71329 & 14.575 &  0.518 &  0.664 &      M\\
 LMC$\_$SC6 & 322363 &  5\uph21\upm54\zdot\ups69 & $-69\arcd23\arcm05\zdot\arcs5$ &  2.34750 &  0.120 &  5.302 &  3.32153 &  0.186 &  4.475 &   0.70675 & 15.052 &  0.595 &  0.630 &      M\\
 LMC$\_$SC8 &  46345 &  5\uph15\upm31\zdot\ups20 & $-69\arcd18\arcm04\zdot\arcs4$ &  2.27345 &  0.145 &  5.435 &  3.17096 &  0.062 &  4.399 &   0.71696 & 15.367 &  0.604 &  0.724 &      M\\
LMC$\_$SC10 & 278863 &  5\uph12\upm00\zdot\ups05 & $-68\arcd48\arcm43\zdot\arcs4$ &  0.97183 &  0.183 &  3.830 &  1.33356 &  0.000 &     -- &   0.72875 & 16.557 &  0.456 &  0.754 &       \\
LMC$\_$SC13 & 156166 &  5\uph06\upm29\zdot\ups53 & $-68\arcd54\arcm20\zdot\arcs2$ &  2.73218 &  0.059 &  5.043 &  3.83507 &  0.168 &  4.363 &   0.71242 & 14.877 &  0.575 &  0.636 &      M\\
LMC$\_$SC14 & 134553 &  5\uph03\upm58\zdot\ups32 & $-69\arcd25\arcm38\zdot\arcs3$ &  2.79677 &  0.049 &  5.749 &  3.89293 &  0.138 &  4.436 &   0.71842 & 14.896 &  0.430 &  0.708 &      M\\
LMC$\_$SC14 & 220934 &  5\uph04\upm29\zdot\ups01 & $-69\arcd09\arcm26\zdot\arcs8$ &  0.86917 &  0.184 &  4.090 &  1.18466 &  0.285 &  4.118 &   0.73369 & 16.406 &  0.361 &  0.543 &       \\
LMC$\_$SC15 &  69667 &  5\uph00\upm55\zdot\ups12 & $-69\arcd16\arcm31\zdot\arcs3$ &  1.37695 &  0.205 &  4.540 &  1.90412 &  0.171 &  4.378 &   0.72314 & 16.052 &  0.564 &  0.691 &      M\\
LMC$\_$SC16 & 115243 &  5\uph35\upm56\zdot\ups77 & $-70\arcd04\arcm51\zdot\arcs2$ &  2.46751 &  0.065 &  5.147 &  3.45510 &  0.193 &  4.441 &   0.71416 & 15.359 &  0.748 &  0.846 &      M\\
LMC$\_$SC16 & 177941 &  5\uph36\upm54\zdot\ups86 & $-70\arcd08\arcm10\zdot\arcs4$ &  1.29934 &  0.080 &  4.464 &  1.78612 &  0.235 &  4.238 &   0.72747 & 16.146 &  0.584 &  0.748 &      M\\
LMC$\_$SC16 & 230285 &  5\uph37\upm38\zdot\ups85 & $-70\arcd14\arcm15\zdot\arcs9$ &  2.55468 &  0.000 &     -- &  3.56455 &  0.142 &  5.164 &   0.71669 & 15.177 &  0.714 &  0.719 &      S\\
LMC$\_$SC17 &  33290 &  5\uph37\upm38\zdot\ups84 & $-70\arcd14\arcm15\zdot\arcs9$ &  2.55469 &  0.034 &  3.970 &  3.56418 &  0.000 &     -- &   0.71677 & 15.174 &  0.706 &  0.711 &       \\
LMC$\_$SC17 & 191865 &  5\uph39\upm29\zdot\ups28 & $-70\arcd38\arcm19\zdot\arcs4$ &  0.93760 &  0.209 &  3.962 &  1.28511 &  0.201 &  4.739 &   0.72959 & 16.436 &  0.933 &  0.642 &      M\\
LMC$\_$SC18 &  89202 &  5\uph41\upm17\zdot\ups55 & $-70\arcd16\arcm38\zdot\arcs7$ &  1.07321 &  0.265 &  4.501 &  1.47140 &  0.263 &  4.161 &   0.72938 & 16.239 &  0.669 &  0.704 &      M\\
\noalign{\vskip2pt}
\hline
\noalign{\vskip2pt}
\multicolumn{15}{l}{Remarks: M: Double-mode Cepheid reported by MACHO (Alcock \etal 1995); S: same star as LMC$\_$SC17 33290}\\
}
\end{landscape}

\begin{landscape}
\renewcommand{\arraystretch}{1.1}
\tabcolsep=5pt
\MakeTable{crccccccccccccl}{12.5cm}{FO/SO Double-Mode Cepheids in the LMC}
{\hline
\noalign{\vskip2pt}
Field   & Star No.& RA(J2000)                & DEC(J2000)                     & $P_{\rm SO}$ & $R_{21}^{\rm SO}$ & $\phi_{21}^{\rm SO}$ & $P_{\rm FO}$ & $R_{21}^{\rm FO}$ & $\phi_{21}^{\rm FO}$ & $P_{\rm SO}/P_{\rm FO}$ & $I$  & ${B-V}$ & ${V-I}$ & Remarks\\
        &         &                          &                                & [days]       &                   &                      & [days]       &                   &                      &                         & [mag]& [mag]   & [mag]   &\\
\noalign{\vskip2pt}
\hline
\noalign{\vskip2pt}
 LMC$\_$SC1 &  44845 &  5\uph32\upm50\zdot\ups01 & $-70\arcd04\arcm15\zdot\arcs1$ &  0.76599 &  0.079 &  5.072 &  0.95199 &  0.227 &  4.202 &  0.80462 & 16.357 &  0.437 &  0.547 &       \\
 LMC$\_$SC1 & 158635 &  5\uph33\upm13\zdot\ups54 & $-69\arcd52\arcm16\zdot\arcs5$ &  0.50726 &  0.106 &  2.885 &  0.62897 &  0.238 &  3.438 &  0.80649 & 17.157 &  0.504 &  0.654 &       \\
 LMC$\_$SC1 & 285275 &  5\uph34\upm34\zdot\ups80 & $-70\arcd18\arcm20\zdot\arcs1$ &  0.68924 &  0.099 &  4.596 &  0.85663 &  0.216 &  4.028 &  0.80459 & 16.515 &  0.467 &  0.608 &      M\\
 LMC$\_$SC1 & 285318 &  5\uph35\upm11\zdot\ups41 & $-70\arcd17\arcm06\zdot\arcs4$ &  0.56838 &  0.129 &  4.194 &  0.70479 &  0.224 &  3.377 &  0.80645 & 16.827 &     -- &     -- &   M,S1\\
 LMC$\_$SC1 & 335559 &  5\uph34\upm32\zdot\ups02 & $-69\arcd45\arcm14\zdot\arcs6$ &  0.60364 &  0.103 &  4.519 &  0.74980 &  0.251 &  3.778 &  0.80507 & 16.727 &  0.371 &  0.645 &      M\\
 LMC$\_$SC2 &  55596 &  5\uph30\upm11\zdot\ups97 & $-69\arcd52\arcm02\zdot\arcs5$ &  0.75139 &  0.000 &     -- &  0.93255 &  0.192 &  4.098 &  0.80574 & 16.419 &  0.440 &  0.626 &      M\\
 LMC$\_$SC2 & 289572 &  5\uph31\upm30\zdot\ups89 & $-69\arcd44\arcm27\zdot\arcs7$ &  0.95258 &  0.000 &     -- &  1.18650 &  0.211 &  4.481 &  0.80285 & 16.282 &  0.426 &  0.743 &       \\
 LMC$\_$SC3 &  53702 &  5\uph27\upm27\zdot\ups70 & $-69\arcd48\arcm07\zdot\arcs9$ &  0.51014 &  0.145 &  2.858 &  0.63296 &  0.209 &  3.295 &  0.80596 & 16.812 &  0.503 &  0.629 &     S2\\
 LMC$\_$SC3 & 187331 &  5\uph28\upm24\zdot\ups13 & $-69\arcd41\arcm55\zdot\arcs1$ &  0.41865 &  0.101 &  3.916 &  0.52051 &  0.177 &  2.949 &  0.80431 & 17.337 &  0.538 &  0.613 &       \\
 LMC$\_$SC3 & 360128 &  5\uph29\upm35\zdot\ups71 & $-70\arcd02\arcm57\zdot\arcs1$ &  0.43604 &  0.280 &  1.653 &  0.54128 &  0.170 &  3.015 &  0.80557 & 16.980 &  0.370 &  0.549 &       \\
 LMC$\_$SC4 &  53796 &  5\uph25\upm26\zdot\ups46 & $-69\arcd49\arcm49\zdot\arcs8$ &  1.03685 &  0.000 &     -- &  1.29506 &  0.158 &  4.294 &  0.80062 & 16.650 &  0.646 &  0.976 &       \\
 LMC$\_$SC4 & 138485 &  5\uph25\upm52\zdot\ups54 & $-70\arcd07\arcm18\zdot\arcs3$ &  0.52044 &  0.000 &     -- &  0.64652 &  0.152 &  3.385 &  0.80499 & 16.806 &  0.423 &  0.516 &       \\
 LMC$\_$SC4 & 168269 &  5\uph26\upm02\zdot\ups18 & $-69\arcd52\arcm10\zdot\arcs0$ &  0.58844 &  0.000 &     -- &  0.72924 &  0.167 &  3.503 &  0.80692 & 16.793 &  0.410 &  0.675 &      M\\
 LMC$\_$SC4 & 176400 &  5\uph25\upm59\zdot\ups29 & $-69\arcd49\arcm14\zdot\arcs2$ &  0.89496 &  0.000 &     -- &  1.10890 &  0.168 &  3.667 &  0.80707 & 16.061 &  0.520 &  0.635 &      M\\
 LMC$\_$SC4 & 220148 &  5\uph26\upm01\zdot\ups36 & $-69\arcd30\arcm41\zdot\arcs7$ &  0.59513 &  0.081 &  4.942 &  0.74085 &  0.247 &  3.650 &  0.80331 & 16.731 &  0.504 &  0.584 &      M\\
 LMC$\_$SC4 & 418294 &  5\uph27\upm27\zdot\ups69 & $-69\arcd48\arcm07\zdot\arcs9$ &  0.51014 &  0.000 &     -- &  0.63296 &  0.202 &  3.267 &  0.80596 & 16.807 &  0.473 &  0.632 &       \\
 LMC$\_$SC5 & 186053 &  5\uph23\upm14\zdot\ups01 & $-69\arcd36\arcm36\zdot\arcs2$ &  0.63112 &  0.138 &  4.634 &  0.78333 &  0.179 &  3.676 &  0.80569 & 16.577 &  0.404 &  0.553 &      M\\
 LMC$\_$SC5 & 338334 &  5\uph23\upm59\zdot\ups19 & $-69\arcd15\arcm29\zdot\arcs6$ &  0.67333 &  0.110 &  4.621 &  0.83483 &  0.184 &  3.548 &  0.80655 & 16.513 &  0.467 &  0.635 &      M\\
 LMC$\_$SC5 & 338399 &  5\uph23\upm49\zdot\ups56 & $-69\arcd13\arcm32\zdot\arcs9$ &  0.46661 &  0.000 &     -- &  0.57951 &  0.121 &  3.550 &  0.80518 & 16.870 &     -- &  0.501 &       \\
 LMC$\_$SC6 &  49297 &  5\uph19\upm57\zdot\ups16 & $-69\arcd39\arcm14\zdot\arcs3$ &  0.41327 &  0.317 &  3.467 &  0.51477 &  0.153 &  4.017 &  0.80282 & 16.913 &     -- &     -- &     S3\\
\noalign{\vskip2pt}
\hline
\noalign{\vskip2pt}
\multicolumn{15}{l}{Remarks: M: Double-mode Cepheid reported by MACHO (Alcock \etal 1999); U: uncertain; S1: same star as LMC$\_$SC16 21126}\\
\multicolumn{15}{l}{~~~~~~~~~~~~~~~S2: same star as LMC$\_$SC4 418294; S3: same star as LMC$\_$SC7 380269}\\
}
\end{landscape}

\begin{landscape}
\renewcommand{\arraystretch}{1.1}
\setcounter{table}{1}
\tabcolsep=5pt
\MakeTable{crccccccccccccl}{12.5cm}{Continued}
{\hline
\noalign{\vskip2pt}
Field   & Star No.& RA(J2000)                & DEC(J2000)                     & $P_{\rm SO}$ & $R_{21}^{\rm SO}$ & $\phi_{21}^{\rm SO}$ & $P_{\rm FO}$ & $R_{21}^{\rm FO}$ & $\phi_{21}^{\rm FO}$ & $P_{\rm SO}/P_{\rm FO}$ & $I$  & ${B-V}$ & ${V-I}$ & Remarks\\
        &         &                          &                                & [days]       &                   &                      & [days]       &                   &                      &                         & [mag]& [mag]   & [mag]   &\\
\noalign{\vskip2pt}
\hline
\noalign{\vskip2pt}
 LMC$\_$SC6 & 142093 &  5\uph21\upm16\zdot\ups55 & $-69\arcd52\arcm02\zdot\arcs6$ &  0.72209 &  0.102 &  4.932 &  0.89629 &  0.184 &  4.061 &  0.80564 & 16.462 &  0.417 &  0.572 &      M\\
 LMC$\_$SC6 & 260869 &  5\uph21\upm25\zdot\ups28 & $-69\arcd52\arcm51\zdot\arcs1$ &  0.65260 &  0.117 &  4.697 &  0.80955 &  0.226 &  3.818 &  0.80613 & 16.761 &  0.443 &  0.676 &      M\\
 LMC$\_$SC6 & 267410 &  5\uph21\upm50\zdot\ups43 & $-69\arcd49\arcm59\zdot\arcs4$ &  0.71666 &  0.235 &  5.142 &  0.88861 &  0.149 &  3.577 &  0.80650 & 16.586 &  0.572 &  0.644 &      U\\
 LMC$\_$SC6 & 413716 &  5\uph22\upm10\zdot\ups24 & $-69\arcd34\arcm18\zdot\arcs1$ &  0.47452 &  0.000 &     -- &  0.59010 &  0.173 &  3.343 &  0.80413 & 17.040 &  0.385 &  0.568 &       \\
 LMC$\_$SC7 & 120511 &  5\uph18\upm10\zdot\ups15 & $-69\arcd50\arcm55\zdot\arcs8$ &  1.00455 &  0.000 &     -- &  1.25127 &  0.189 &  4.090 &  0.80282 & 15.947 &  0.494 &  0.590 &       \\
 LMC$\_$SC7 & 207275 &  5\uph18\upm42\zdot\ups46 & $-69\arcd11\arcm15\zdot\arcs2$ &  0.50097 &  0.088 &  4.970 &  0.62125 &  0.199 &  3.175 &  0.80639 & 16.925 &  0.506 &  0.579 &       \\
 LMC$\_$SC7 & 221814 &  5\uph18\upm28\zdot\ups01 & $-69\arcd03\arcm25\zdot\arcs7$ &  0.32917 &  0.000 &     -- &  0.40999 &  0.101 &  2.906 &  0.80287 & 17.564 &  0.433 &  0.517 &       \\
 LMC$\_$SC7 & 380269 &  5\uph19\upm57\zdot\ups15 & $-69\arcd39\arcm14\zdot\arcs4$ &  0.41338 &  0.107 &  0.175 &  0.51477 &  0.140 &  3.410 &  0.80304 & 16.845 &  0.322 &  0.518 &       \\
 LMC$\_$SC7 & 425296 &  5\uph20\upm02\zdot\ups63 & $-69\arcd23\arcm54\zdot\arcs3$ &  0.32400 &  0.000 &     -- &  0.40354 &  0.172 &  2.676 &  0.80289 & 17.607 &  0.506 &  0.525 &      U\\
 LMC$\_$SC8 &    142 &  5\uph15\upm35\zdot\ups86 & $-69\arcd45\arcm48\zdot\arcs3$ &  0.75174 &  0.078 &  4.792 &  0.93471 &  0.250 &  4.061 &  0.80425 & 16.403 &  0.664 &  0.540 &       \\
 LMC$\_$SC8 &  10158 &  5\uph15\upm06\zdot\ups51 & $-69\arcd39\arcm52\zdot\arcs9$ &  0.55567 &  0.094 &  4.761 &  0.69000 &  0.235 &  3.406 &  0.80532 & 16.421 &  0.706 &  0.684 &   M,S4\\
 LMC$\_$SC8 &  81678 &  5\uph15\upm35\zdot\ups57 & $-68\arcd57\arcm07\zdot\arcs4$ &  0.53637 &  0.110 &  4.088 &  0.66520 &  0.214 &  3.166 &  0.80633 & 16.918 &  0.662 &  0.591 &      M\\
 LMC$\_$SC8 & 198932 &  5\uph16\upm28\zdot\ups69 & $-69\arcd36\arcm33\zdot\arcs1$ &  0.57415 &  0.141 &  4.672 &  0.71155 &  0.199 &  3.310 &  0.80690 & 16.819 &  0.509 &  0.582 &      M\\
 LMC$\_$SC8 & 218854 &  5\uph16\upm28\zdot\ups51 & $-69\arcd25\arcm35\zdot\arcs7$ &  0.60137 &  0.200 &  4.018 &  0.74787 &  0.205 &  3.655 &  0.80411 & 16.859 &  0.671 &  0.685 &      M\\
 LMC$\_$SC8 & 242825 &  5\uph16\upm55\zdot\ups71 & $-69\arcd08\arcm49\zdot\arcs7$ &  0.92936 &  0.090 &  4.392 &  1.15428 &  0.240 &  4.463 &  0.80514 & 16.132 &  0.460 &  0.598 &       \\
 LMC$\_$SC8 & 326147 &  5\uph17\upm18\zdot\ups24 & $-69\arcd16\arcm38\zdot\arcs9$ &  0.82110 &  0.070 &  5.502 &  1.02239 &  0.231 &  4.151 &  0.80312 & 16.303 &  0.466 &  0.619 &       \\
 LMC$\_$SC8 & 337664 &  5\uph17\upm20\zdot\ups66 & $-69\arcd09\arcm29\zdot\arcs1$ &  0.50728 &  0.000 &     -- &  0.62979 &  0.171 &  3.493 &  0.80547 & 16.718 &  0.480 &  0.480 &       \\
 LMC$\_$SC9 & 270100 &  5\uph14\upm16\zdot\ups93 & $-68\arcd54\arcm13\zdot\arcs4$ &  0.40602 &  0.169 &  4.448 &  0.50487 &  0.258 &  3.261 &  0.80421 & 17.321 &     -- &  0.596 &       \\
 LMC$\_$SC9 & 286128 &  5\uph15\upm06\zdot\ups52 & $-69\arcd39\arcm52\zdot\arcs9$ &  0.55567 &  0.286 &  4.988 &  0.69000 &  0.288 &  3.515 &  0.80532 & 16.415 &  0.573 &  0.715 &      M\\
LMC$\_$SC10 & 204083 &  5\uph11\upm39\zdot\ups97 & $-68\arcd49\arcm57\zdot\arcs6$ &  0.42292 &  0.000 &     -- &  0.52626 &  0.163 &  3.416 &  0.80363 & 17.202 &     -- &     -- &      M\\
\noalign{\vskip2pt}
\hline
\noalign{\vskip2pt}
\multicolumn{15}{l}{Remarks: M: Double-mode Cepheid reported by MACHO (Alcock \etal 1999); U: uncertain; S4: same star as LMC$\_$SC9 286128}\\
}
\end{landscape}

\begin{landscape}
\renewcommand{\arraystretch}{1.1}
\setcounter{table}{1}
\tabcolsep=5pt
\MakeTable{crccccccccccccl}{12.5cm}{Continued}
{\hline
\noalign{\vskip2pt}
Field   & Star No.& RA(J2000)                & DEC(J2000)                     & $P_{\rm SO}$ & $R_{21}^{\rm SO}$ & $\phi_{21}^{\rm SO}$ & $P_{\rm FO}$ & $R_{21}^{\rm FO}$ & $\phi_{21}^{\rm FO}$ & $P_{\rm SO}/P_{\rm FO}$ & $I$  & ${B-V}$ & ${V-I}$ & Remarks\\
        &         &                          &                                & [days]       &                   &                      & [days]       &                   &                      &                         & [mag]& [mag]   & [mag]   &\\
\noalign{\vskip2pt}
\hline
\noalign{\vskip2pt}
LMC$\_$SC11 &  38029 &  5\uph07\upm36\zdot\ups83 & $-69\arcd12\arcm46\zdot\arcs1$ &  0.97544 &  0.000 &     -- &  1.21809 &  0.191 &  4.185 &  0.80079 & 16.182 &  0.533 &  0.672 &      M\\
LMC$\_$SC11 & 130342 &  5\uph08\upm21\zdot\ups30 & $-69\arcd07\arcm17\zdot\arcs5$ &  0.53341 &  0.218 &  4.108 &  0.66087 &  0.198 &  3.498 &  0.80713 & 16.979 &  0.487 &  0.578 &       \\
LMC$\_$SC11 & 186270 &  5\uph09\upm07\zdot\ups10 & $-69\arcd29\arcm21\zdot\arcs4$ &  0.38792 &  0.000 &     -- &  0.48332 &  0.234 &  3.124 &  0.80262 & 17.523 &  0.514 &  0.640 &       \\
LMC$\_$SC11 & 233290 &  5\uph09\upm08\zdot\ups12 & $-68\arcd56\arcm42\zdot\arcs9$ &  0.97835 &  0.095 &  4.702 &  1.21750 &  0.167 &  4.050 &  0.80357 & 16.022 &  0.519 &  0.646 &      M\\
LMC$\_$SC15 &  16385 &  5\uph00\upm24\zdot\ups15 & $-69\arcd14\arcm57\zdot\arcs1$ &  0.79571 &  0.078 &  0.845 &  0.99044 &  0.253 &  4.164 &  0.80339 & 16.265 &  0.392 &  0.544 &       \\
LMC$\_$SC15 &  85604 &  5\uph00\upm54\zdot\ups79 & $-69\arcd03\arcm42\zdot\arcs0$ &  0.51857 &  0.000 &     -- &  0.64280 &  0.131 &  3.464 &  0.80674 & 16.953 &  0.324 &  0.608 &       \\
LMC$\_$SC15 &  96008 &  5\uph00\upm38\zdot\ups77 & $-68\arcd53\arcm55\zdot\arcs5$ &  0.79736 &  0.117 &  1.998 &  0.99170 &  0.243 &  4.002 &  0.80403 & 16.381 &  0.473 &  0.596 &       \\
LMC$\_$SC15 & 152481 &  5\uph01\upm48\zdot\ups72 & $-68\arcd50\arcm44\zdot\arcs6$ &  0.89088 &  0.000 &     -- &  1.11021 &  0.269 &  4.147 &  0.80244 & 16.041 &  0.479 &  0.572 &      U\\
LMC$\_$SC15 & 208026 &  5\uph02\upm09\zdot\ups94 & $-68\arcd51\arcm31\zdot\arcs1$ &  1.05956 &  0.064 &  3.742 &  1.32122 &  0.210 &  3.831 &  0.80196 & 15.755 &  0.566 &  0.693 &      M\\
LMC$\_$SC16 &  21126 &  5\uph35\upm11\zdot\ups41 & $-70\arcd17\arcm06\zdot\arcs6$ &  0.56842 &  0.255 &  3.651 &  0.70478 &  0.190 &  3.378 &  0.80652 & 16.818 &  0.521 &  0.628 &      M\\
LMC$\_$SC16 &  31616 &  5\uph35\upm27\zdot\ups74 & $-70\arcd12\arcm13\zdot\arcs4$ &  0.63788 &  0.132 &  4.873 &  0.79140 &  0.211 &  3.811 &  0.80601 & 16.907 &  0.562 &  0.717 &       \\
LMC$\_$SC16 &  37231 &  5\uph35\upm33\zdot\ups45 & $-70\arcd08\arcm41\zdot\arcs5$ &  0.59719 &  0.000 &     -- &  0.74026 &  0.208 &  3.383 &  0.80673 & 16.766 &  0.521 &  0.631 &       \\
LMC$\_$SC16 & 266808 &  5\uph37\upm36\zdot\ups40 & $-69\arcd44\arcm20\zdot\arcs4$ &  1.08041 &  0.178 &  3.603 &  1.35292 &  0.151 &  4.464 &  0.79858 & 16.571 &  0.994 &  1.094 &      M\\
LMC$\_$SC17 &  80292 &  5\uph38\upm45\zdot\ups56 & $-70\arcd36\arcm11\zdot\arcs5$ &  0.65024 &  0.000 &     -- &  0.80782 &  0.228 &  3.657 &  0.80493 & 16.518 &  0.436 &  0.607 &      M\\
LMC$\_$SC17 & 186042 &  5\uph38\upm51\zdot\ups24 & $-69\arcd49\arcm21\zdot\arcs8$ &  0.49119 &  0.140 &  2.244 &  0.61099 &  0.206 &  3.513 &  0.80392 & 17.084 &     -- &  0.669 &      M\\
LMC$\_$SC18 & 199230 &  5\uph42\upm13\zdot\ups76 & $-70\arcd09\arcm13\zdot\arcs5$ &  0.59250 &  0.000 &     -- &  0.73438 &  0.210 &  3.509 &  0.80680 & 17.068 &  0.627 &  0.860 &       \\
LMC$\_$SC20 &  21200 &  5\uph45\upm22\zdot\ups28 & $-70\arcd50\arcm11\zdot\arcs1$ &  0.89659 &  0.141 &  4.917 &  1.11954 &  0.227 &  4.170 &  0.80086 & 16.346 &  0.618 &  0.717 &      M\\
LMC$\_$SC20 & 138333 &  5\uph46\upm43\zdot\ups89 & $-70\arcd40\arcm51\zdot\arcs4$ &  0.69215 &  0.000 &     -- &  0.85985 &  0.241 &  4.193 &  0.80497 & 16.588 &  0.377 &  0.623 &      M\\
LMC$\_$SC20 & 188572 &  5\uph47\upm12\zdot\ups66 & $-70\arcd41\arcm13\zdot\arcs1$ &  0.81482 &  0.000 &     -- &  1.01544 &  0.212 &  4.007 &  0.80243 & 16.350 &  0.622 &  0.714 &      M\\
LMC$\_$SC21 &  12012 &  5\uph20\upm19\zdot\ups62 & $-70\arcd42\arcm29\zdot\arcs0$ &  1.07489 &  0.087 &  5.251 &  1.34153 &  0.189 &  4.197 &  0.80124 & 15.833 &  0.557 &  0.607 &      M\\
LMC$\_$SC21 & 178950 &  5\uph22\upm06\zdot\ups50 & $-70\arcd16\arcm11\zdot\arcs6$ &  0.86768 &  0.113 &  5.490 &  1.08130 &  0.220 &  4.289 &  0.80244 & 16.193 &  0.483 &  0.578 &       \\
\noalign{\vskip2pt}
\hline
\noalign{\vskip2pt}
\multicolumn{15}{l}{Remarks: M: Double-mode Cepheid reported by MACHO (Alcock \etal 1999); U: uncertain}\\
}
\end{landscape}

\Section{Double-Mode Cepheids in the LMC}
\vspace*{-4pt}
Tables~1 and 2 list all double-mode Cepheids detected in the central
area  of the LMC. They contain 81 entries but only 76 objects. Five
stars are located in the overlapping regions between fields  and they
were discovered independently in each field. Fundamental and  first
overtone mode pulsators (FU/FO) are listed in Table~1 while  Table~2
includes objects pulsating in the first and second overtones  (FO/SO).
Basic parameters of each star: right ascension and declination  (J2000),
the intensity-mean {\it I}-band magnitude, ${(B-V)}$ and ${(V- I)}$
colors, both periods and their ratio are provided. Accuracy of  periods
is about ${4\cdot10^{-5}P}$. Finding charts for all objects are 
presented in Appendix~A. The size of the {\it I}-band subframes is 
${60\times60}$ arcsec; North is up and East to the left. 

Appendices~B and C show the light curves of FU/FO and FO/SO pulsators, 
respectively. The first and second columns in each Appendix contain 
original photometric data folded with the shorter and longer periods 
while the remaining columns show variability attributed to each mode 
after subtraction of the other period variability approximated by 
Fourier series of fifth order. For objects revealing also periodicity 
equal to the sum and/or difference of both mode frequencies and having 
an amplitude larger than twice the formal error -- such terms were also 
subtracted from the original data. {\it BVI} photometry of all objects 
is available from the OGLE Internet archive (see Section~6). 

Completeness of the sample is determined by completeness of the variable star 
search in the OGLE databases and efficiency of double-mode Cepheid detection 
algorithm. OGLE Cepheid catalog of the LMC was estimated to be complete in 
more than 96\% (Udalski \etal 1999d). Completeness of the detection algorithm 
can be assessed by comparison of results obtained in the preliminary and final 
({\sc Clean}) searches. More than 90\% objects in both lists are common 
suggesting good completeness of the search. 

As a test of completeness we cross-identified double mode Cepheids 
reported by MACHO (Alcock \etal 1995, 1999). 39 out of 42 objects which 
are located in the OGLE fields were detected during our search. Two of 
the remaining objects have been misclassified, and have not 
entered to the list of Cepheids. One star -- LMC\_SC4 168269 -- has a 
faint, close companion, and the noise in the power spectrum was greater 
than the peak of its second pulsation mode. All double-mode Cepheids found 
by the MACHO team are marked by the letter 'M' in the last column of 
Tables~1 and 2. 
\vspace*{-6pt}
\Section{Discussion}
\vspace*{-4pt}
Seventy six double-mode Cepheids were identified during the presented 
search in the 4.5 square degree area in the central part of the LMC. 
Nineteen objects pulsate simultaneously in the fundamental mode and 
first overtone while 57 objects in the first and second overtones. In 
the following Subsections we present the basic observational properties 
of detected sample. 

\Subsection{Period Ratio in Double-Mode Cepheids}
The ratio of the two periods of double-mode Cepheids is expected to be 
dependent on metallicity. This dependence is clearly seen in Fig.~1, 
which presents the period ratio of the FU/FO pulsators plotted as a 
function of the fundamental mode period. Beside of the LMC double-mode 
Cepheids we present there the period ratios of FU/FO Cepheids from the 
SMC (Udalski \etal 1999a) and the Galactic FU/FO Cepheids (Pardo and 
Poretti 1997). The metal contents of Cepheids in the Galaxy, LMC  and
SMC are approximately ${Z=0.02}$, 0.008 and 0.004, respectively. It  can
be seen, that the lower metallicity -- the higher the FU/FO period ratio
of double-mode Cepheids. The best linear fit for the LMC FU/FO pulsators is
given  by the equation: 
\begin{eqnarray}
P_{\rm FO}/P_{\rm FU} = 0.734 & - & 0.035\times \log P_{\rm FU},\\
0.002 & & 0.004\nonumber
\end{eqnarray}
\begin{figure}[htb]
\includegraphics[width=12.5cm, bb=25 45 505 405]{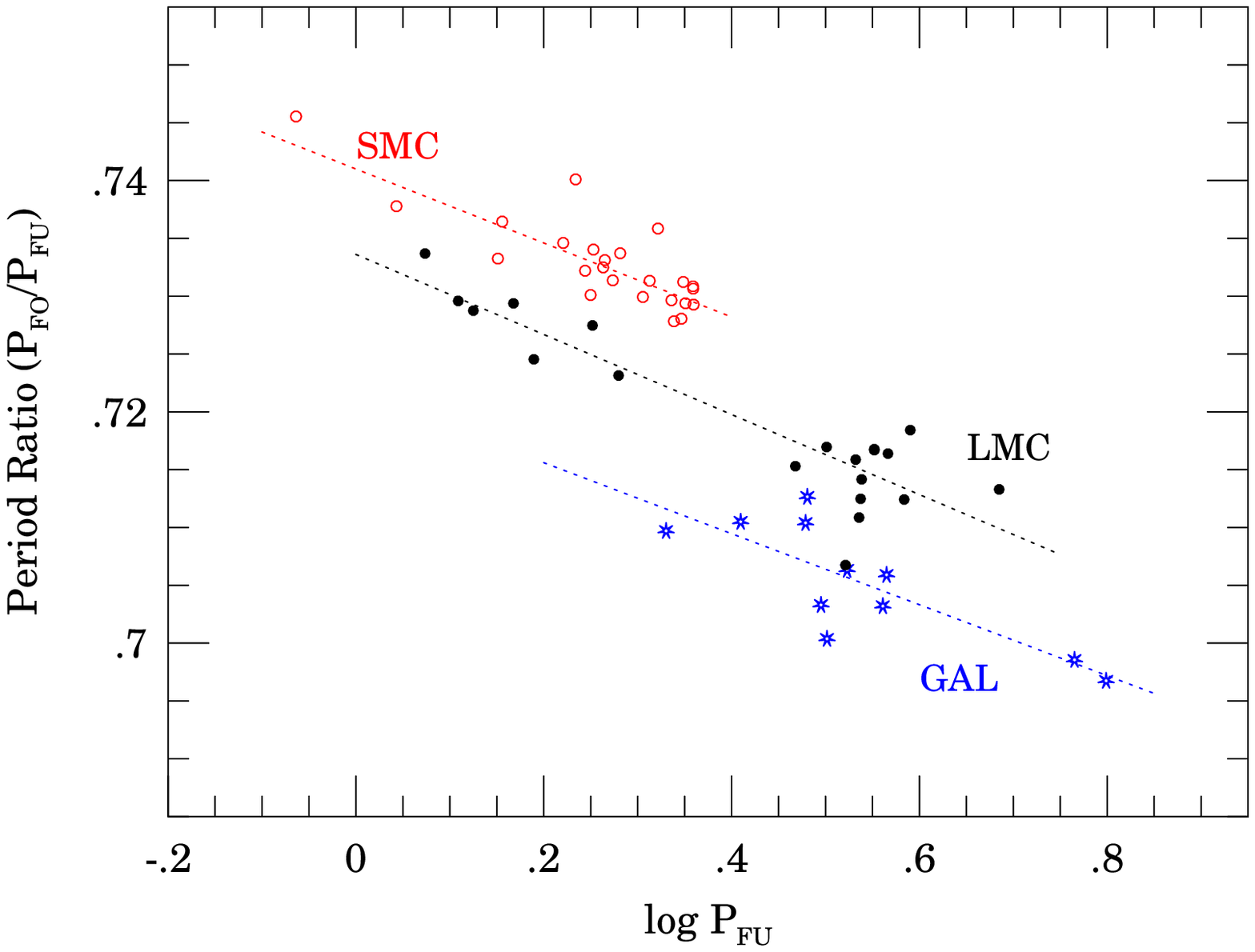}
\FigCap{Ratio of periods in FU/FO double mode Cepheids in the Galaxy
and Magellanic Clouds. Dotted lines indicate the best linear fits to the
observed ratios.}
\end{figure} 

Period ratios for FO/SO Cepheids in LMC are almost identical with the 
ratios of the SMC objects. Fig.~2 presents these values in similar 
fashion as Fig.~1. The only one known FO/SO Galactic Cepheid (CO~Aur) is 
also marked on the diagram. Our sample of the LMC double-mode Cepheids is 
numerous enough to note that the period ratio-period relation for FO/SO 
Cepheids is non-linear. The best square fit is as follows: 
$$P_{\rm SO}/P_{\rm FO}=0.804-0.020\times\log P_{\rm FO}-
0.067\times\log^2P_{\rm FO}.\eqno(2)$$
\begin{figure}[htb]
\includegraphics[width=12.5cm, bb=24 45 505 405]{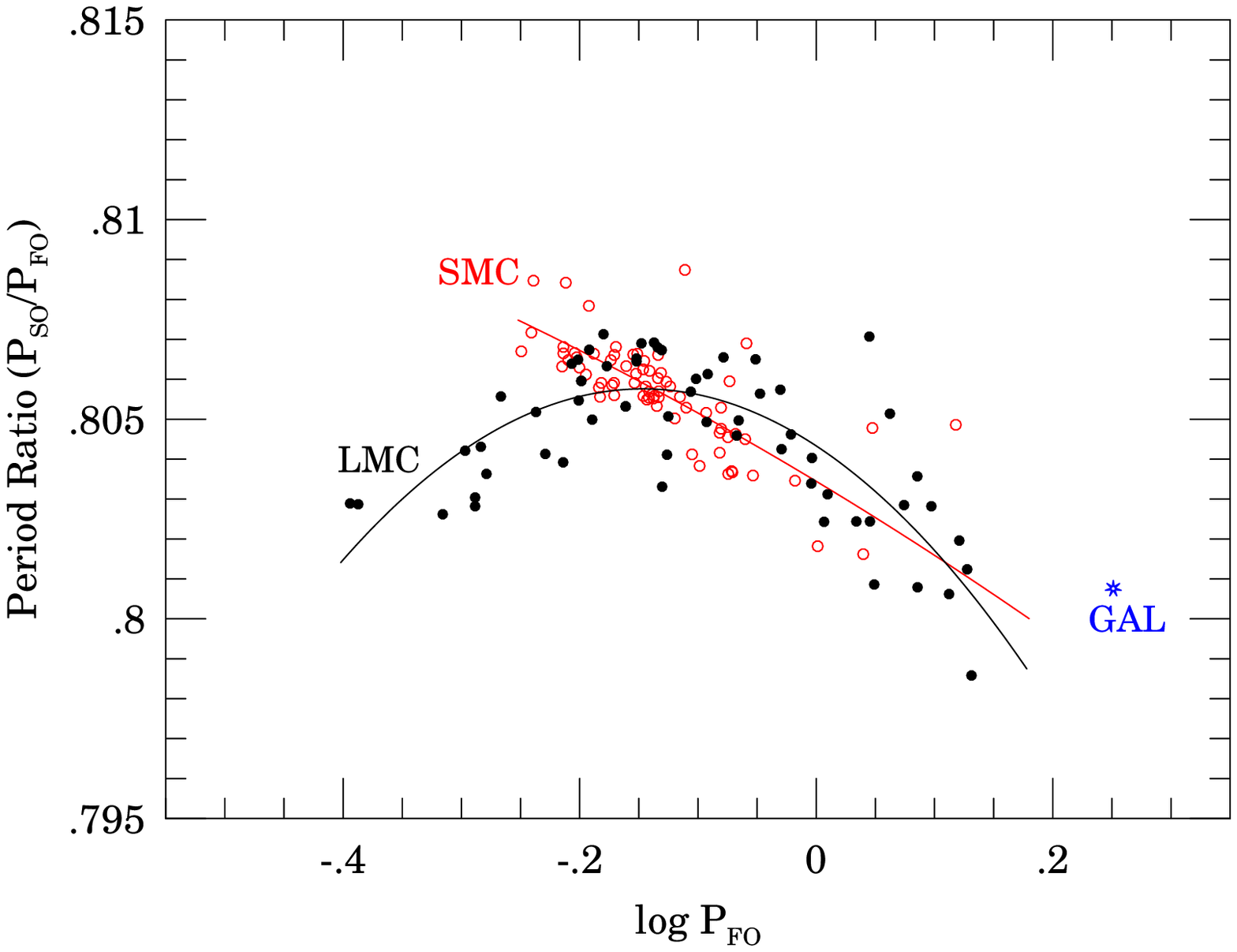}
\FigCap{Ratio of periods in FO/SO double mode Cepheids in the Galaxy and 
Magellanic Clouds. Thick solid line indicates the best square fit to the LMC 
data (filled dots) while thin line to the SMC data (open dots).} 
\end{figure} 

\Subsection{Fourier Decomposition of Light Curves of Double-Mode Cepheids}
Fourier decomposition of light curves of pulsating stars has been widely 
used for analyzing their properties (Simon and Lee 1981). In the case of 
Cepheids the ratio of amplitudes of the first harmonic and the 
fundamental period, ${R_{21}=A_2/A_1}$, and phase difference, 
${\phi_{21}=\phi_2-2\phi_1}$ are particularly useful. Both allow to 
distinguish between fundamental mode and first overtone pulsators. 

Fig.~3 presents the $R_{21}$ \vs $\log P$ and $\phi_{21}$ \vs $\log P$ 
diagrams constructed for more than 1300 single-mode Cepheids (small 
dots) taken from the Catalog of Cepheids from the LMC (Udalski \etal 
1999d). The $R_{21}$ \vs $\log P$ diagram shows the characteristic, well 
separated "V-shape" sequences for Cepheids pulsating in the fundamental 
mode and the first overtone. In the similar diagram $\phi_{21}$ \vs 
$\log P$ the sequences for both modes of pulsation are also well defined 
but the separation is smaller and in some ranges of periods they 
overlap. 

\begin{figure}[p]
\hglue-.7cm{\includegraphics[width=13.5cm, bb=25 40 505 710]{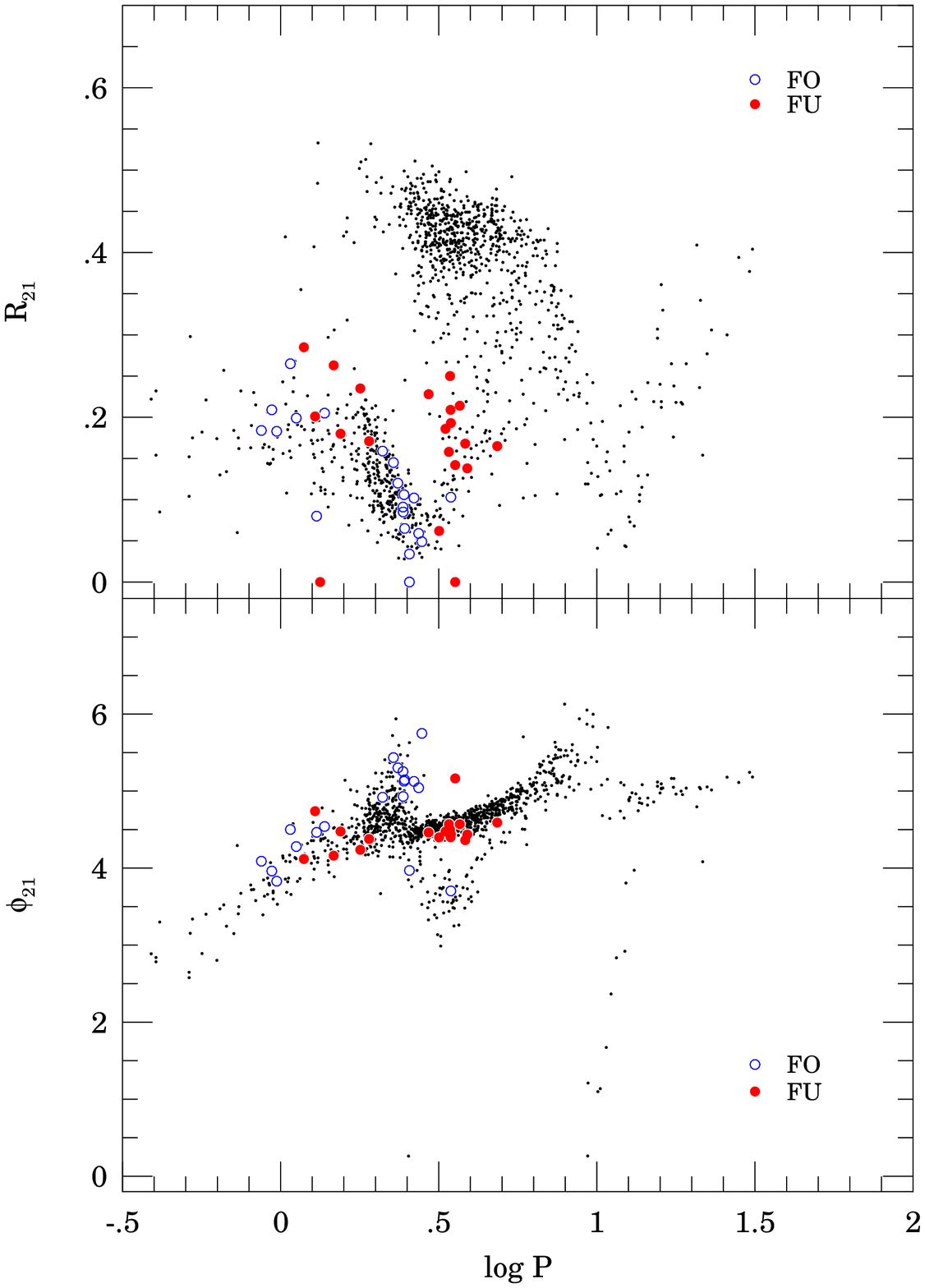}}
\FigCap{${R_{21}}$ and ${\phi_{21}}$ \vs $\log P$ diagrams for
single-mode Cepheids from the LMC (small dots). Large open and filled
circles mark values of the first overtone and fundamental mode
pulsations in the FU/FO double-mode Cepheids, respectively.}
\end{figure} 
\begin{figure}[p]
\hglue-.7cm{\includegraphics[width=13.5cm, bb=25 40 505 710]{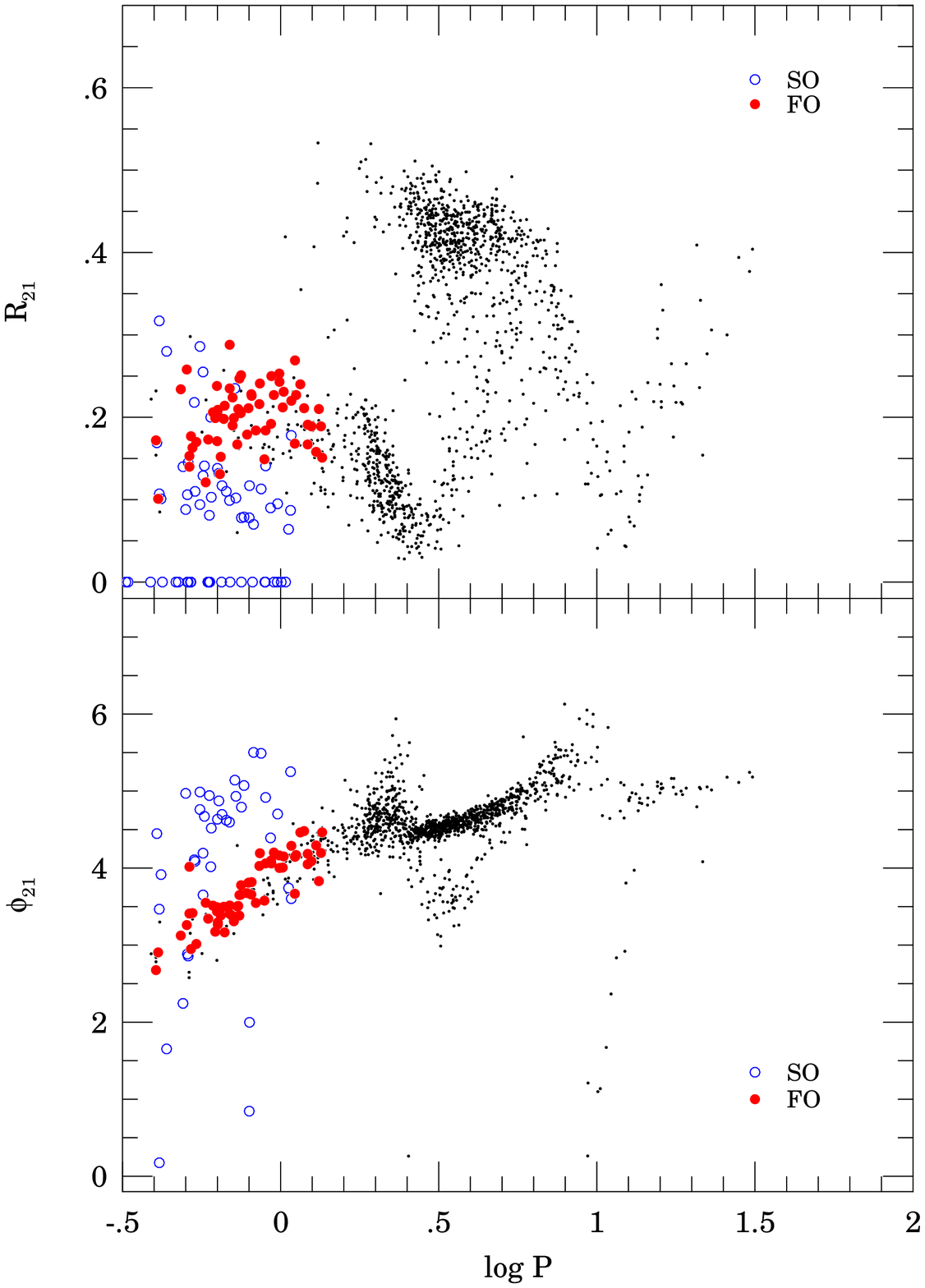}}
\FigCap{${R_{21}}$ and ${\phi_{21}}$ \vs $\log P$ diagrams for
single-mode Cepheids from the LMC (small dots). Large open and filled
circles mark values of the second and first overtone pulsations in the
FO/SO double-mode Cepheids, respectively.}
\end{figure} 
We decomposed the light curves of double-mode Cepheids to the sum of two 
Fourier series of fifth order corresponding to both periodicities 
including the terms with periodicities equal to the sum and difference of 
mode frequencies when their amplitudes were larger than twice the formal 
errors. Then we calculated $R_{21}$ and $\phi_{21}$ for both pulsating 
modes. They are listed in Tables~1 and 2. 

Fig.~3 shows positions of the FU/FO Cepheids on the $R_{21}$ \vs $\log
P$  and $\phi_{21}$ \vs $\log P$ diagrams. The values for the
fundamental  mode pulsation are plotted with large filled dots while for
the first  overtone mode with open circles. Objects with non-significant
first  harmonic amplitude, $A_2$, (\ie with almost sinusoidal light
curve) have  ${R_{21}=0}$ and their $\phi_{21}$ is not defined. Similar
diagrams for  FO/SO double-mode Cepheids are shown in Fig.~4. Values of
$R_{21}$ and $\phi_{21}$ for  the first overtone  pulsation are plotted
with large  filled dots and with open circles for the second overtone
pulsations.

Figs.~3 and 4 are similar to analogous diagrams plotted for the SMC 
double-mode Cepheids (Udalski \etal 1999a). In both types of double-mode 
Cepheids the values of $R_{21}$ and $\phi_{21}$ of first overtone pulsations 
fall on the sequences of the single-mode first overtone Cepheids. 

Fundamental mode pulsations in the FU/FO Cepheids have much smaller values of 
the $R_{21}$ parameter compared to the single-mode fundamental mode pulsators. 
They fall practically on the sequence of the first overtone  pulsators. This 
means that the light curves of the fundamental mode  pulsations in double-mode 
Cepheids are more sinusoidal than in the single-mode Cepheids of that type. 
However, it is clearly seen that $\phi_{21}$ of the fundamental mode 
pulsations in double-mode Cepheids fall in  most cases on the single-mode 
fundamental mode Cepheid sequence. 

Second overtone pulsations in the FO/SO Cepheids are generally small
amplitude, almost sinusoidal variations (small ${R_{21}}$). Their
$\phi_{21}$ values are usually larger than the first overtone values for
similar periodicities (Fig.~4). Identical conclusions on the behavior of
second overtone pulsations in double-mode Ceheids were presented by
Alcock \etal (1999).

\Subsection{Color-Magnitude Diagram and Colors of Double-Mode Cepheids}
Fig.~5 presents the color-magnitude diagram of subfield 2 of the LMC\_SC3 
field corrected for the mean $E(B-V)=0.120$ reddening in this direction 
(Udalski \etal 1999d). Tiny dots correspond to the field stars. Larger 
dots indicate positions of the single-mode Cepheids: fundamental mode -- 
darker dots, first overtone -- lighter dots. Positions of FU/FO 
double-mode Cepheids are indicated by large filled circles while FO/SO 
objects by star symbols. 
\begin{figure}[p]
\hglue-.9cm{\includegraphics[width=13cm, bb=25 40 505 710]{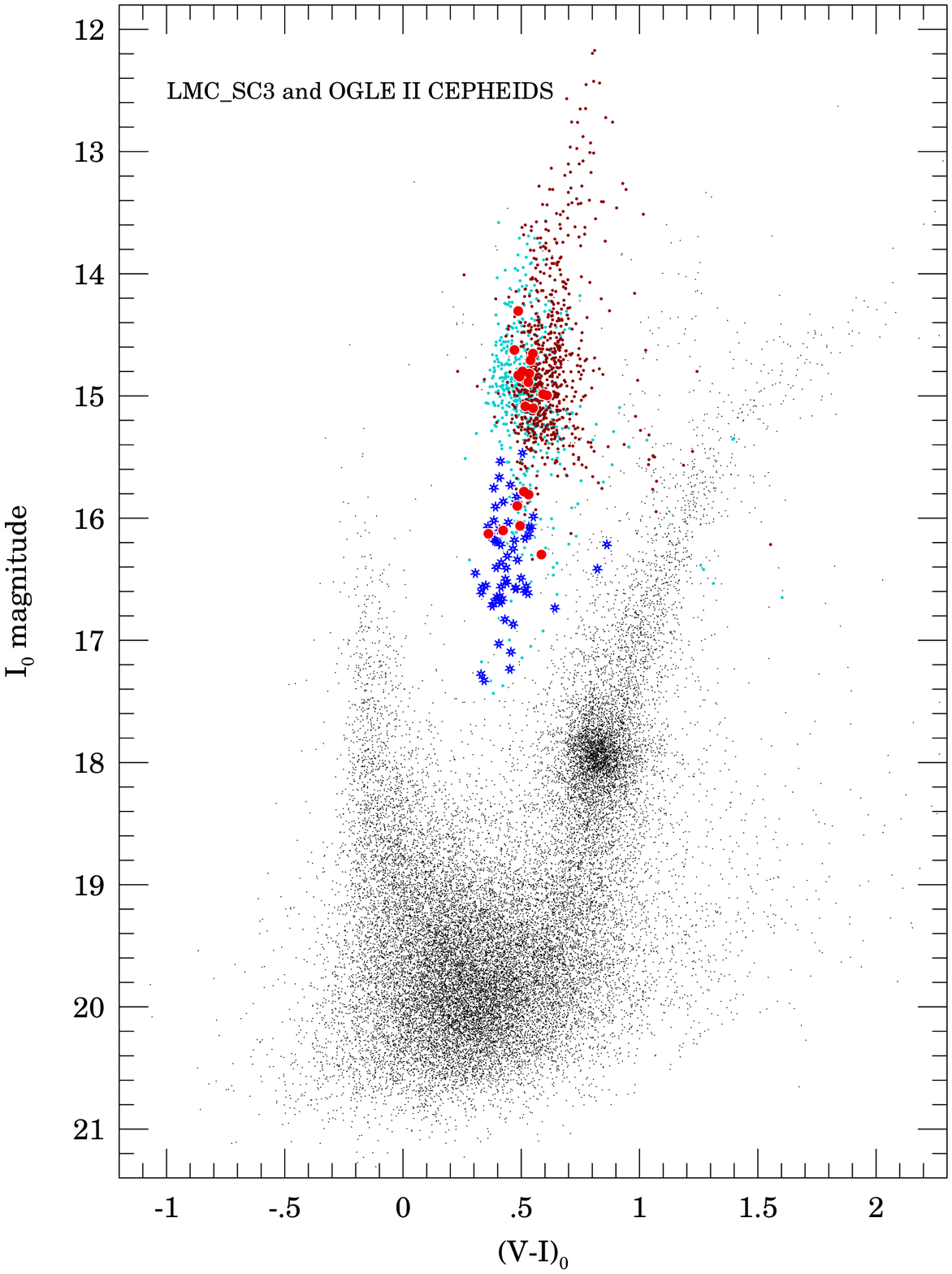}}
\FigCap{Color-magnitude diagram of the LMC$\_$SC3 field. Only about 8\%
of the field stars are plotted by tiny dots. Larger dots show positions of
single-mode  fundamental type Cepheids (darker dots) and first overtone
stars (lighter  dots). Large filled circles and star symbols mark
positions of the FU/FO and FO/SO double-mode Cepheids, respectively.}
\end{figure} 

\begin{figure}[htb]
\hglue-9mm{\includegraphics[width=13.5cm, bb=25 45 505 405]{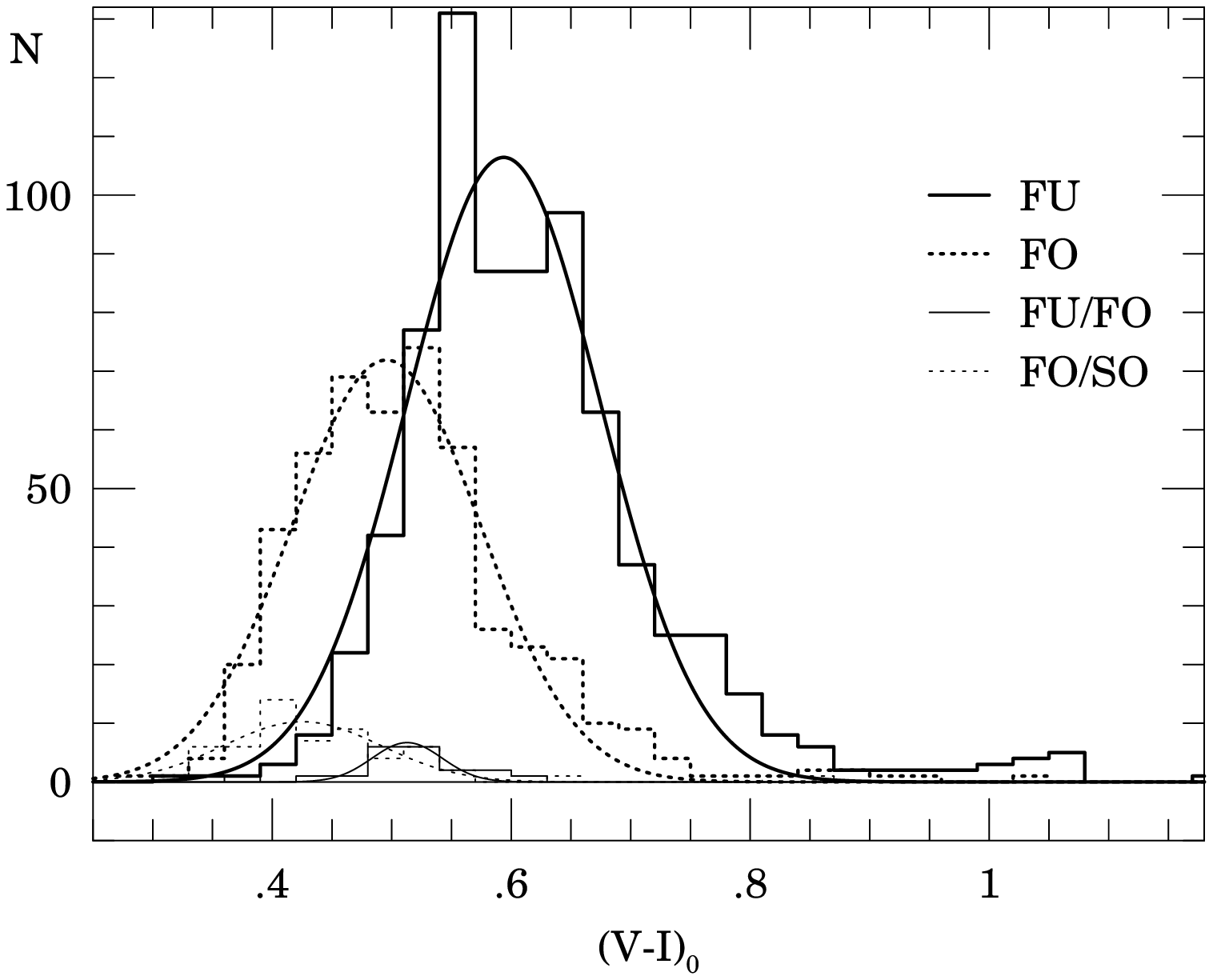}}
\FigCap{Histograms of $(V-I)_0$ color distribution of single-mode and
double-mode Cepheids in the LMC. Thick lines represent distribution of
single-mode Cepheids: solid line -- fundamental mode pulsators, dotted
line -- first overtone objects. Distribution of double-mode Cepheids is
marked by thin line: solid line -- FU/FO stars, dotted line -- FO/SO
Cepheids. The bins are 0.03~mag wide.}
\end{figure} 
Fig.~6 shows distribution of color indices $(V-I)_0$ of single-mode FU and FO 
Cepheids and double-mode FU/FO and FO/SO pulsators in the LMC. ${E(B-V)}$ 
reddening values from Table~2 of Udalski \etal (1999d) were used to deredden 
the photometry presented in Tables~1 and 2. The width of the bin is 0.03~mag. 
Thick solid and dotted lines correspond to the single-mode fundamental and 
first overtone Cepheids while thin solid and dotted lines to the FU/FO and 
FO/SO double-mode pulsators. All histograms were fitted with Gaussians which 
fit well the observed color distributions. In the case of single-mode 
fundamental mode Cepheids there is a small excess of red objects. 

The mean $(V-I)_0$ color and the standard deviation of its distribution 
are: (0.425, 0.07), (0.496, 0.08), (0.513, 0.03), (0.594, 0.08) 
respectively for FO/SO, FO, FU/FO and FU Cepheids. Color indices 
(temperature) distribution depends on the type of pulsations. The 
single-mode first overtone Cepheids are on average by about 0.1~mag 
bluer than the fundamental mode pulsators. As one could expect the FU/FO 
double-mode Cepheids have ${V-I}$ color distribution in between the 
first and fundamental mode distributions of single-mode stars. The color 
distribution of FO/SO double-mode Cepheids resembles that of the 
single-mode first overtone stars but it is shifted bluewards. 

\Section{Data Availability}
The {\it BVI} photometry of the LMC double-mode Cepheids is available to
the astronomical community in the electronic form from the OGLE archive: 
\begin{center}
{\it http://www.astrouw.edu.pl/\~{}ogle} \\
{\it ftp://sirius.astrouw.edu.pl/ogle/ogle2/var\_stars/lmc/cep/dmcep/}\\
\end{center}
or its US mirror
\begin{center}
{\it http://bulge.princeton.edu/\~{}ogle}\\
{\it ftp://bulge.princeton.edu/ogle/ogle2//var\_stars/lmc/cep/dmcep/}\\
\end{center}

\Acknow{The paper was partly supported by the Polish KBN grant 2P03D00814 to 
A.\ Udalski and 2P03D00916 to M.\ Szyma{\'n}ski. Partial support for the OGLE 
project was provided with the NSF grant AST-9820314 to B.~Paczy\'nski.}

\end{document}